\documentclass[11pt,twoside]{article}
\usepackage{FUSE2004}
\usepackage{natbib}

\usepackage{epsf}
\usepackage{psfig}
\usepackage{lscape}

\begin{document}
\title{Metals in the Neutral Interstellar Medium of Dwarf Star-Forming Galaxies}
\author{A.~Aloisi$^{1,2}$, T.~M.~Heckman$^{3}$, C.~G.~Hoopes$^{3}$, C.~Leitherer$^{1}$,
S.~Savaglio$^{3}$, K.~R.~Sembach$^{1}$}
\affil{
$^1$\,Space Telescope Science Institute, 3700 San Martin Drive, Baltimore, 
MD 21218\newline
$^2$\,On assignment from the Space Telescope Division of ESA\newline
$^3$\,Departmnent of Physics and Astronomy, Johns Hopkins University, 
3400 North Charles Street, Baltimore, MD 21218
}

\begin{abstract}
The determination of the metal abundances in the neutral interstellar
medium (ISM) of dwarf star-forming galaxies is a key step in understanding
their physical and chemical evolution. This type of investigation has been 
possible in the last 5 years thanks to FUSE. We will give a flavor of the 
issues involved by presenting the work that we are performing in this 
astrophysical field.\looseness=-2
\end{abstract}

\section{Introduction}

Dwarf galaxies are the most common type of galaxies in the local 
Universe and may have also played an important role in the past 
as building blocks from which larger systems formed by merging. 
Present-day dwarfs are the leftovers of this building process, 
and as such they could have had a completely different evolutionary 
path compared to their high-reshift counterparts. However, local 
star-forming dwarfs are particularly interesting from a cosmological 
point of view. They are chemically poorly evolved systems which 
are undergoing star formation (SF), thus the closest analogue to 
primeval galaxies. They are also perfect targets for FUSE, since 
SF means hot young and massive stars that emit the bulk of their 
energy in the rest-frame UV rich in transitions of the most abundant 
ions. These galaxies are characterized by a large reservoir of 
H\,{\sc i}. FUSE is thus particularly suited to study the physics, 
chemistry and kinematics of their ISM.\looseness=-2

\section{FUSE Spectra of I~Zw~18}

I~Zw~18 is the star-forming galaxy with the lowest metallicity 
known and has always been regarded as the best candidate for 
a truly ``young'' galaxy in the local universe. It has been 
observed with FUSE for a total of $\sim 90$ (60) ksec in the 
LiF (SiC) channels. The LWRS aperture (30''$\times\,$30'') was 
used to cover the whole body for a resolution of $\sim 35$ km/s 
and a S/N of 7-18 (Aloisi et al.~2003). A multi-component fitting 
technique was applied to infer the column densities of the most 
common heavy elements and determine metal abundances in the 
neutral gas. Our results are reported in Table~1 together with 
the metal abundances in the H\,{\sc ii} regions (Izotov et 
al.~1999; Izotov \& Thuan 1999).

\begin{table}[!ht]
\caption{Interstellar Abundances in I~Zw~18}
\smallskip
\begin{center}
{\small
\begin{tabular}{llcccc}
\tableline
\noalign{\smallskip}
Element & Ion & & [X/H]$_{\rm ISM}$ & & [X/H]$_{\rm HII}$\\[3pt] 
\noalign{\smallskip}
\tableline
\noalign{\smallskip}
O  & O~{\sc i}   & & $-2.06 \pm 0.28$ & & $-1.51 \pm 0.04$\\  
Ar & Ar~{\sc i}  & & $-2.27 \pm 0.13$ & & $-1.51 \pm 0.07$\\  
Si & Si~{\sc ii} & & $-2.09 \pm 0.12$ & & $-1.90 \pm 0.33$\\  
N  & N~{\sc i}   & & $-2.88 \pm 0.11$ & & $-2.36 \pm 0.07$\\ 
Fe & Fe~{\sc ii} & & $-1.76 \pm 0.12$ & & $-1.96 \pm 0.09$\\
\noalign{\smallskip}
\tableline\
\end{tabular}
}
\end{center}
\end{table}

It is clear that the neutral gas in I~Zw~18 has already been 
enriched in heavy elements, and is not primordial in nature. 
The $\alpha$ elements are several times lower in the H\,{\sc i} 
than in the H\,{\sc ii} gas, while Fe is the same. The Fe behavior 
suggests that some old SF is required for the metal enrichment of 
the H\,{\sc i} (Fe is mostly produced by SNe\,Ia on time 
scales $> 1$ Gyr). The relative metal content in $\alpha$ elements
(produced by SNe\,II on time scales $< 50$ Myr) and N (released 
on timescales $> 300$ Myr) between neutral and ionized gas suggests 
that the H\,{\sc ii} regions have been additionally enriched by 
more recent SF.

\section{FUSE and STIS Spectra of NGC~1705}

There are some caveats to address in order to be sure that what 
we really measure are the ISM metal abundances. First of all, there
is saturation, especially for O\,{\sc i}. Hidden saturation of 
unresolved multiple components can also bring to erroneous estimates 
of the column density. Ionization constitutes another type of 
uncertainty. Abundances are derived by assuming that the primary 
ionization state in the neutral gas is representative of the total 
amount of a certain element and this is probably the case in ISM 
studies. However, we can have contamination by ionized gas laying 
along the line of sight. Furthermore, Ar\,{\sc i} and N\,{\sc i} 
should be well coupled with H\,{\sc i} and O\,{\sc i} due to similar 
ionization potentials. However, they could be found in higher 
percentage in their ionized state due to larger cross-sections for 
photoionization. Finally, depletion is important. Some elements 
could be more easily locked into dust grains than others (e.g., 
Fe compared to O), thus altering the relative abundances produced 
by a certain SF history.\looseness=-2

A wonderful dataset where to address all these issues 
is represented by the FUSE and STIS Echelle 
spectra (900-3100 \AA) of NGC~1705, one of the brightest dwarf 
starburst galaxies in the nearby Universe. The FUSE data were 
taken by centering the SSC in the LWRS aperture for a total of 
$\sim 21$ ksec, a resolution of $\sim 30$ km/s and a S/N of 
10-16 (Heckman et al.~2001). The STIS 
echelle data were taken with the 0.2''$\times\,$0.2'' aperture 
centered on the SSC for a total of 10 HST orbits, a resolution of 
$\sim 15$ km/s, and a S/N of 10-20 
(V\'azquez et al.~2004). We measured the column density of 
many ions in the FUSE and STIS spectra of NGC~1705 with the 
line-profile fitting and inferred the ISM metal abundances. We 
found consistency in the measurements performed on those ions 
detected in both spectra. Thus, the FUV light is 
dominated by the SSC. The low-ionization absorption lines have a 
mean radial velocity of about 590 km/s. However, two components 
were detected for selected ions in the higher-resolution STIS 
data. One component is at the same radial velocity of the stars 
in the SSC ($v = 618$ km/s) as inferred by the stellar C\,{\sc iii} 
line at 1175 \AA, and the second at the velocity of the warm photoionized 
gas ($v = 580$ km/s) as detected in absorption through C\,{\sc iii} 
or N\,{\sc ii} and confirmed by nebular emission lines in the optical. 
The use of the abundances from the total (neutral $+$ ionized) absorbing 
gas would thus be misleading for the derivation of the metal content in 
the neutral ISM of NGC~1705. In Table~2 we report the abundances as 
inferred by both the total column densities of the ions (column 3) and 
the column densities of the absorbing component at rest with the stars 
in the SSC (column 4). 
The latter have to be compared with the H\,{\sc ii} region abundances
(Lee \& Skillman 2004).\looseness=-2

\begin{table}[!ht]
\caption{Interstellar Abundances in NGC~1705}
\smallskip
\begin{center}
{\small
\begin{tabular}{llcccccc}
\tableline
\noalign{\smallskip}
Element & Ion & & [X/H]$_{\rm ISM,total}$ & & [X/H]$_{\rm ISM,neutral}$ & 
& [X/H]$_{\rm HII}$\\[3pt] 
\noalign{\smallskip}
\tableline
\noalign{\smallskip}
O  & O~{\sc i}   & & $-1.19 \pm 0.01$ & &        ...       & & $-0.48 \pm 0.05$\\  
Ar & Ar~{\sc i}  & & $-1.11 \pm 0.04$ & &        ...       & & $-0.61 \pm 0.10$\\  
Si & Si~{\sc ii} & & $-0.90 \pm 0.01$ & &        ...       & &        ...      \\  
Mg & Mg~{\sc ii} & & $-1.41 \pm 0.12$ & &        ...       & &        ...      \\ 
Al & Al~{\sc ii} & & $-1.14 \pm 0.04$ & & $-1.36 \pm 0.05$ & &        ...      \\ 
N  & N~{\sc i}   & & $-1.79 \pm 0.03$ & & $-2.29 \pm 0.06$ & & $-1.51 \pm 0.08$\\
Fe & Fe~{\sc ii} & & $-0.86 \pm 0.03$ & & $-1.29 \pm 0.03$ & &        ...      \\
\noalign{\smallskip}
\tableline\
\end{tabular}
}
\end{center}
\end{table}

\section{Conclusions}

The offset in metal content between neutral ISM and H\,{\sc ii} regions 
in dwarf star-forming galaxies is probably one of the unexpected great 
results of FUSE in the past five years. However, this area of research 
is still pretty new and many unknowns and uncertainties still affect the 
interpretation of the data. It is thus premature to draw conclusions and 
more targets with a data quality similar to that of NGC~1705, still need 
to be investigated before having all the pieces of this intriguing puzzle 
put together.






\end{document}